# Equivalence Classes in AES – Part 1

David Cornwell: dave.cornwell@yahoo.com

**Abstract:** We investigate properties of equivalence classes in AES which arise naturally from properties of MixColumns and InvMixColumns. These two operations have the property that the XOR of the 4 input bytes equals the XOR of 4 output bytes. We examine the effect on equivalence classes due to the operation of SubBytes, ShiftRows, MixColumns and AddRoundKey. The next phase of research is to find a key recovery attack using known (plaintext, ciphertext) equivalence class pairs.

## Introduction

AES has been standardized for 25 years and is used to secure the internet, personal communications, intranets of corporations and military networks, to secure data from Unclassified to Top Secret. Therefore, any weakness in this cryptographic algorithm is vitally important for all concerned.

Cryptanalysis is a science as well as an art. It requires both theoretical knowledge and experimentation to discover properties and weaknesses of cryptographic algorithms. Quite often it is easier to discover a new property using experimentation rather than theoretical analysis and later justify it theoretically. In the following, we report on results that were discovered using both theoretical and experimental methods. In this paper, we only consider AES-128, but the ideas are also applicable to AES-192 and AES-256.

In the following, + indicates a bitwise XOR operation and multiplication is over $GF(2^8)$ using $m(x) = x^8 + x^4 + x^3 + x + 1$.

## What is an equivalence class?

The round structure of AES uses the following operations on a 4x4 8-bit byte state:

a) SubBytes
b) ShiftRows
c) MixColumns
d) AddRoundKey

The MixColumns operation applies the following matrix to a column of 4 bytes:

$$\begin{pmatrix} 2 & 3 & 1 & 1 \\ 1 & 2 & 3 & 1 \\ 1 & 1 & 2 & 3 \\ 3 & 1 & 1 & 2 \end{pmatrix}$$

Property 1a: MixColumns preserves the XOR of the four input bytes

Proof:

Let $(X1, X2, X3, X4)$ be the four input bytes to MixColumns and $(Y1, Y2, Y3, Y4)$ be the four output bytes. Then we have the following where the + operation is byte-wise XOR, and multiplication is modulo m(x):

$Y1 = 2X1 + 3X2 + X3 + X4$

$Y2 = X1 + 2X2 + 3X3 + X4$

$Y3 = X1 + X2 + 2X3 + 3X4$

$Y4 = 3X1 + X2 + X3 + 2X4$

Consider

$Y1 + Y2 + Y3 + Y4 = (2 + 1 + 1 + 3)X1 + (3 + 2 + 1 + 1)X2 + (1 + 3 + 2 + 1)X3 + (1 + 1 + 3 + 2)X4 = X1 + X2 + X3 + X4$

because addition is over GF[2] and $2 + 3 + 1 + 1 = 1$.

QED

The InvMixColumns operation applies the following matrix to a column of 4 bytes:

$$\begin{pmatrix} 14 & 11 & 13 & 9 \\ 9 & 14 & 11 & 13 \\ 13 & 9 & 14 & 11 \\ 11 & 13 & 9 & 14 \end{pmatrix}$$

Property 1b: InvMixColumns preserves the XOR of the four input bytes

Proof:

Let $(X1, X2, X3, X4)$ be the four input bytes to InvMixColumns and $(Y1, Y2, Y3, Y4)$ be the four output bytes. Then we have the following where the + operation is byte-wise XOR,

$Y1 = 14X1 + 11X2 + 13X3 + 9X4$

$Y2 = 9X1 + 14X2 + 11X3 + 13X4$

$Y3 = 13X1 + 9X2 + 14X3 + 11X4$

$Y4 = 11X1 + 13X2 + 9X3 + 14X4$

Consider,

$Y1 + Y2 + Y3 + Y4 = (14 + 9 + 13 + 11)X1 + (11 + 14 + 9 + 13)X2 + (13 + 11 + 14 + 9)X3 + (9 + 13 + 11 + 14)X4 = X1 + X2 + X3 + X4$

because addition is over GF[2] and $14 + 9 + 13 + 11 = 1110 + 1001 + 1101 + 1011\ binary = 1$.
QED.

There are 256 possible XOR sums of four input bytes {0,1,2, ...,255} which are preserved in the sum of the output bytes using MixColumns and InvMixColumns.

Definition: An equivalence class E(Q) is the set of $2^{24}$ 4-byte vectors $(X1, X2, X3, X4)$ whose 8-bit XOR is the fixed 8-bit value Q. i.e. Q= X1+X2+X3+X4, where + is the XOR operation. We call Q the equivalence class value.

In the following sections we examine the composition of two operations, MixColumns(ShiftRows(S)) applied to a state S, and the effect on equivalence class values, in the forward (encryption) direction. We also examine the effect of InvShiftRows(InvMixColumns(S)) applied to a state S in the backward (decryption) direction.

We show that if we know the 32-bit equivalence class values (Q1, Q2, Q3, Q4) of the 128-bit input state S, we can compute the 32-bit equivalence class values (R1, R2, R3, R4) of the output state, after applying either MixColumns(ShiftRows(S)) or InvShiftRows(InvMixColumns(S)), by simply computing appropriate linear combinations of the input equivalence class values. In other words, to compute (R1, R2, R3, R4) we do not have to know the full 128-bit state S, we just need to know (Q1, Q2, Q3, Q4) and a corresponding 4x4 matrix:

$$\begin{pmatrix} q11 & q12 & q13 & q14 \\ q21 & q22 & q23 & q24 \\ q31 & q32 & q33 & q34 \\ q41 & q42 & q43 & q44 \end{pmatrix} \begin{pmatrix} Q1 \\ Q2 \\ Q3 \\ Q4 \end{pmatrix} = \begin{pmatrix} R1 \\ R2 \\ R3 \\ R4 \end{pmatrix}$$

## Effect of ShiftRows followed by MixColumns Applied to State S1

Property 2a (One Shift):

Let the 128-bit state of AES be S1 which has the property that each of the four 32-bit columns belong to E(Q1), E(Q2), E(Q3) and E(Q4) respectively:

$$S1 = \begin{pmatrix} A11 & A12 & A13 & A14 \\ A21 & A22 & A23 & A24 \\ A31 & A32 & A33 & A34 \\ A41 & A42 & A43 & A44 \end{pmatrix}$$

That is, the XOR of each column is Q1, Q2, Q3 and Q4 respectively. We map the 128-bit state down to the 32-bit state vector (Q1, Q2, Q3, Q4). S1 is then shifted by ShiftRows once to get the state S2:

$$S2 = \begin{pmatrix} A11 & A12 & A13 & A14 \\ A22 & A23 & A24 & A21 \\ A33 & A34 & A31 & A32 \\ A44 & A41 & A42 & A43 \end{pmatrix}$$

then we apply MixColumns to each column of S2 to get the state S3:

$$S3 = \begin{pmatrix} B11 & B12 & B13 & B14 \\ B22 & B23 & B24 & B21 \\ B33 & B34 & B31 & B32 \\ B44 & B41 & B42 & B43 \end{pmatrix}$$

Then we have,

A) $B11 + B21 + B31 + B41 = 2Q1 + 3Q2 + Q3 + Q4 = R1$
B) $B12 + B22 + B32 + B42 = Q1 + 2Q2 + 3Q3 + Q4 = R2$
C) $B13 + B23 + B33 + B43 = Q1 + Q2 + 2Q3 + 3Q4 = R3$
D) $B14 + B24 + B34 + B44 = 3Q1 + Q2 + Q3 + 2Q4 = R4$

Rewriting this we have,

$$\begin{pmatrix} 2 & 3 & 1 & 1 \\ 1 & 2 & 3 & 1 \\ 1 & 1 & 2 & 3 \\ 3 & 1 & 1 & 2 \end{pmatrix} \begin{pmatrix} Q1 \\ Q2 \\ Q3 \\ Q4 \end{pmatrix} = \begin{pmatrix} R1 \\ R2 \\ R3 \\ R4 \end{pmatrix}$$

Proof:

Consider the sum of $B11 + B21 + B31 + B41$ in S3. We have,

$B11 = 2A11 + 3A22 + A33 + A44$

$B21 = A14 + 2A21 + 3A32 + A43$

$B31 = A13 + A24 + 2A31 + 3A42$

$B41 = 3A12 + A23 + A34 + 2A41$

Then let $R1 = B11 + B21 + B31 + B41$, we have

$R1 = 2(A11 + A21 + A31 + A41) + 3(A22 + A32 + A42 + A12) + (A33 + A43 + A13 + A23) + (A44 + A14 + A24 + A34)$

Therefore, since

$A11 + A21 + A31 + A41 = Q1$

$A22 + A32 + A42 + A12 = Q2$

$A33 + A43 + A13 + A23 = Q3$

$A44 + A14 + A24 + A34 = Q4$

we have

$R1 = B11 + B21 + B31 + B41 = 2Q1 + 3Q2 + Q3 + Q4$

Then we have shown condition A) of Property 2a. The results for the conditions B), C) and D) can be found in a similar fashion.

QED

## Effect of InvMixColumns followed by InvShiftRows Applied to State S3
<u>Property 2b (One Shift)</u>:

Let the 128-bit state of AES be S3 which has the property that each of the four 32-bit shifted columns belong to E(R1), E(R2), E(R3) and E(R4) respectively:

$$S3 = \begin{pmatrix} B11 & B12 & B13 & B14 \\ B22 & B23 & B24 & B21 \\ B33 & B34 & B31 & B32 \\ B44 & B41 & B42 & B43 \end{pmatrix}$$

That is, the XOR of each shifted column is R1, R2, R3 and R4 respectively. We map the 128-bit state down to the 32-bit state vector (R1, R2, R3, R4). S3 has InvMixColumns applied to it to get the state S2, followed by InvShiftRows to get the state S1 i.e. we apply the inverse operations of property 2a in the corresponding reverse order:

$$S2 = \begin{pmatrix} A11 & A12 & A13 & A14 \\ A22 & A23 & A24 & A21 \\ A33 & A34 & A31 & A32 \\ A44 & A41 & A42 & A43 \end{pmatrix}$$

$$S1 = \begin{pmatrix} A11 & A12 & A13 & A14 \\ A21 & A22 & A23 & A24 \\ A31 & A32 & A33 & A34 \\ A41 & A42 & A43 & A44 \end{pmatrix}$$

Then we have,

A) $A11 + A21 + A31 + A41 = 14R1 + 11R2 + 13R3 + 9R4 = Q1$
B) $A12 + A22 + A32 + A42 = 9R1 + 14R2 + 11R3 + 13R4 = Q2$
C) $A13 + A23 + A33 + A43 = 13R1 + 9R2 + 14R3 + 11R4 = Q3$
D) $A14 + A24 + A34 + A44 = 11R1 + 13R2 + 9R3 + 14R4 = Q4$

Rewriting this we have,

$$\begin{pmatrix} 14 & 11 & 13 & 9 \\ 9 & 14 & 11 & 13 \\ 13 & 9 & 14 & 11 \\ 11 & 13 & 9 & 14 \end{pmatrix} \begin{pmatrix} R1 \\ R2 \\ R3 \\ R4 \end{pmatrix} = \begin{pmatrix} Q1 \\ Q2 \\ Q3 \\ Q4 \end{pmatrix}$$

Poof:

Consider the sum of $A11 + A21 + A31 + A41$ in S2. We have,

$A11 = 14B11 + 11B22 + 13B33 + 9B44$

$A21 = 9B14 + 14B21 + 11B32 + 13B43$

$A31 = 13B13 + 9B24 + 14B31 + 11B42$

$A41 = 11B12 + 13B23 + 9B34 + 14B41$

Then let $Q1 = A11 + A21 + A31 + A41$, we have

$Q1 = 14(B11 + B21 + B31 + B41) + 11(B22 + B32 + B42 + B12)$
$\qquad + 13(B33 + B43 + B13 + B23) + 9(B44 + B14 + B24 + B34)$

Therefore, since

$B11 + B21 + B31 + B41 = R1$

$B22 + B32 + B42 + B12 = R2$

$B33 + B43 + B13 + B23 = R3$

$B44 + B14 + B24 + B34 = R4$

we have

$Q1 = A11 + A21 + A31 + A41 = 14R1 + 11R2 + 13R3 + 9R4$

Then we have shown condition A) of Property 2b. The results for the conditions B), C) and D) can be found in a similar fashion.

QED

## Effect of ShiftRows followed by MixColumns Applied to State S3

<u>Property 3a (Two Shifts):</u>

Let the 128-bit state of AES be S3 from Property 2a, which has the property that each of the four shifted 32-bit columns belong to E(R1), E(R2), E(R3) and E(R4) respectively, as shown in Property 2a:

$$S3 = \begin{pmatrix} B11 & B12 & B13 & B14 \\ B22 & B23 & B24 & B21 \\ B33 & B34 & B31 & B32 \\ B44 & B41 & B42 & B43 \end{pmatrix}$$

That is, the XOR of each MIxColumns(ShiftRows(S1)) shifted column is R1, R2, R3 and R4 respectively. We map the 128-bit state down to the 32-bit state vector (R1, R2, R3, R4). S3 is then shifted by ShiftRows once to get the state S4:

$$S4 = \begin{pmatrix} B11 & B12 & B13 & B14 \\ B23 & B24 & B21 & B22 \\ B31 & B32 & B33 & B34 \\ B43 & B44 & B41 & B42 \end{pmatrix}$$

then we apply MixColumns to each column of S4 to get the state S5:

$$S5 = \begin{pmatrix} C11 & C12 & C13 & C14 \\ C23 & C24 & C21 & C22 \\ C31 & C32 & C33 & C34 \\ C43 & C44 & C41 & C42 \end{pmatrix}$$

Then we have,

A) $C11 + C21 + C31 + C41 = 3R1 + 2R3 = T1$
B) $C12 + C22 + C32 + C42 = 3R2 + 2R4 = T2$
C) $C13 + C23 + C33 + C43 = 2R1 + 3R3 = T3$
D) $C14 + C24 + C34 + C44 = 2R2 + 3R4 = T4$

Rewriting this we have,

$$\begin{pmatrix} 3 & 0 & 2 & 0 \\ 0 & 3 & 0 & 2 \\ 2 & 0 & 3 & 0 \\ 0 & 2 & 0 & 3 \end{pmatrix} \begin{pmatrix} R1 \\ R2 \\ R3 \\ R4 \end{pmatrix} = \begin{pmatrix} T1 \\ T2 \\ T3 \\ T4 \end{pmatrix}$$

Proof:

Consider the sum of $T1 = C11 + C21 + C31 + C41$ in S5. We have,

$C11 = 2B11 + 3B23 + B31 + B43$

$C21 = B13 + 2B21 + 3B33 + B41$

$C31 = B11 + B23 + 2B31 + 3B43$

$C41 = 3B13 + B21 + B33 + 2B41$

Then we have,

$T1 = 2(B11 + B21 + B31 + B41) + 3(B23 + B33 + B43 + B13) + (B31 + B41 + B11 + B21) + (B43 + B13 + B23 + B33)$

Therefore, since

$B11 + B21 + B31 + B41 = R1$

$B33 + B43 + B13 + B23 = R3$

we have

$T1 = C11 + C21 + C31 + C41 = 3R1 + 2R3$

Then we have shown condition A) of Property 3. The results for the conditions B), C) and D) can be found in a similar fashion.

QED

## Effect of InvMixColumns followed by InvShiftRows Applied to State S5

Property 3b (Two Shifts):

Let the 128-bit state of AES be S5 from Property 3a, which has the property that each of the four shifted 32-bit columns belong to E(T1), E(T2), E(T3) and E(T4) respectively, as shown in Property 3a:

$$S5 = \begin{pmatrix} C11 & C12 & C13 & C14 \\ C23 & C24 & C21 & C22 \\ C31 & C32 & C33 & C34 \\ C43 & C44 & C41 & C42 \end{pmatrix}$$

That is, the XOR of each shifted column of S5 is T1, T2, T3 and T4 respectively. We map the 128-bit state down to the 32-bit state vector (T1, T2, T3, T4). S5 is then applied by InvMixColumns to get the state S4:

$$S4 = \begin{pmatrix} B11 & B12 & B13 & B14 \\ B23 & B24 & B21 & B22 \\ B31 & B32 & B33 & B34 \\ B43 & B44 & B41 & B42 \end{pmatrix}$$

then we apply InvShiftRows to S4 to get the state S3:

$$S3 = \begin{pmatrix} B11 & B12 & B13 & B14 \\ B22 & B23 & B24 & B21 \\ B33 & B34 & B31 & B32 \\ B44 & B41 & B42 & B43 \end{pmatrix}$$

Then we have,

A) $B11 + B21 + B31 + B41 = 3T1 + 2T3 = R1$
B) $B12 + B22 + B32 + B42 = 3T2 + 2T4 = R2$
C) $B13 + B23 + B33 + B43 = 2T1 + 3T3 = R3$
D) $B14 + B24 + B34 + B44 = 2T2 + 3T4 = R4$

Rewriting this we have,

$$\begin{pmatrix} 3 & 0 & 2 & 0 \\ 0 & 3 & 0 & 2 \\ 2 & 0 & 3 & 0 \\ 0 & 2 & 0 & 3 \end{pmatrix} \begin{pmatrix} T1 \\ T2 \\ T3 \\ T4 \end{pmatrix} = \begin{pmatrix} R1 \\ R2 \\ R3 \\ R4 \end{pmatrix}$$

Proof:

Consider the sum of $R1 = B11 + B21 + B31 + B41$ in S4. We have,

$B11 = 14C11 + 11C23 + 13C31 + 9C43$

$B21 = 9C13 + 14C21 + 11C33 + 13C41$

$B31 = 13C11 + 9C23 + 14C31 + 11C43$

$B41 = 11C13 + 13C21 + 9C33 + 14C41$

Then we have,

$R1 = 14(C11 + C21 + C31 + C41) + 11(C13 + C23 + C33 + C43)$
$\quad + 13(C11 + C21 + C31 + C41) + 9(C13 + C23 + C33 + C43)$

Therefore, since

$C11 + C21 + C31 + C41 = T1$

$C13 + C23 + C33 + C43 = T3$

we have

$R1 = B11 + B21 + B31 + B41 = 3T1 + 2T3$

Then we have shown condition A) of Property 3b. The results for the conditions B), C) and D) can be found in a similar fashion.

QED

# Effect of ShiftRows followed by MixColumns Applied to State S5

Property 4a (Three Shifts):

Let the 128-bit state of AES be S5 from Property 3a, which is such that each of the four shifted 32-bit columns belong to E(T1), E(T2), E(T3) and E(T4) respectively, as shown in Property 3a:

$$S5 = \begin{pmatrix} C11 & C12 & C13 & C14 \\ C23 & C24 & C21 & C22 \\ C31 & C32 & C33 & C34 \\ C43 & C44 & C41 & C42 \end{pmatrix}$$

That is, the XOR of each MIxColumns(ShiftRows(S3)) shifted columns are T1, T2, T3 and T4 respectively. We map the 128-bit state down to the 32-bit state vector (T1, T2, T3, T4). S5 is then shifted by ShiftRows once to get the state S6:

$$S6 = \begin{pmatrix} C11 & C12 & C13 & C14 \\ C24 & C21 & C22 & C23 \\ C33 & C34 & C31 & C32 \\ C42 & C43 & C44 & C41 \end{pmatrix}$$

then we apply MixColumns to each column of S6 to get the state S7:

$$S7 = \begin{pmatrix} D11 & D12 & D13 & D14 \\ D24 & D21 & D22 & D23 \\ D33 & D34 & D31 & D32 \\ D42 & D43 & D44 & D41 \end{pmatrix}$$

Then we have,

A) $D11 + D21 + D31 + D41 = 2T1 + T2 + T3 + 3T4 = U1$
B) $D12 + D22 + D32 + D42 = 3T1 + 2T2 + T3 + T4 = U2$
C) $D13 + D23 + D33 + D43 = T1 + 3T2 + 2T3 + T4 = U3$
D) $D14 + D24 + D34 + D44 = T1 + T2 + 3T3 + 2T4 = U4$

Rewriting this we have.

$$\begin{pmatrix} 2 & 1 & 1 & 3 \\ 3 & 2 & 1 & 1 \\ 1 & 3 & 2 & 1 \\ 1 & 1 & 3 & 2 \end{pmatrix} \begin{pmatrix} T1 \\ T2 \\ T3 \\ T4 \end{pmatrix} = \begin{pmatrix} U1 \\ U2 \\ U3 \\ U4 \end{pmatrix}$$

Proof:

Consider the sum of $U1 = D11 + D21 + D31 + D41$ in S7. We have,

$D11 = 2C11 + 3C24 + C33 + C42$

$D21 = C12 + 2C21 + 3C34 + C43$

$D31 = C13 + C22 + 2C31 + 3C44$

$D41 = 3C14 + C23 + C32 + 2C41$

Then we have,

$$U1 = 2(C11 + C21 + C31 + C41) + 3(C24 + C34 + C44 + C14) + (C33 + C43 + C13 + C23) + (C42 + C12 + C22 + C32)$$

Therefore, since

$$C11 + C21 + C31 + C41 = T1$$

$$C12 + C22 + C32 + C42 = T2$$

$$C13 + C23 + C33 + C43 = T3$$

$$C14 + C24 + C34 + C44 = T4$$

we have

$$U1 = D11 + D21 + D31 + D41 = 2T1 + T2 + T3 + 3T4$$

Then we have shown condition A) of Property 4a. The results for the conditions B), C) and D) can be found in a similar fashion.

QED

# Effect of InvMixColumns followed by InvShiftRows Applied to State S7

Property 4b (Three Shifts):

Let the 128-bit state of AES be S7 from Property 4a, which is such that each of the four shifted 32-bit columns belong to E(U1), E(U2), E(U3) and E(U4) respectively, as shown in Property 4a:

$$S7 = \begin{pmatrix} D11 & D12 & D13 & D14 \\ D24 & D21 & D22 & D23 \\ D33 & D34 & D31 & D32 \\ D42 & D43 & D44 & D41 \end{pmatrix}$$

That is, the XOR of each shifted column of S7 are U1, U2, U3 and U4 respectively. We map the 128-bit state down to the 32-bit state vector (U1, U2, U3, U4). S7 then has InvMixColumns applied to it to get S6:

$$S6 = \begin{pmatrix} C11 & C12 & C13 & C14 \\ C24 & C21 & C22 & C23 \\ C33 & C34 & C31 & C32 \\ C42 & C43 & C44 & C41 \end{pmatrix}$$

then, we apply InvShiftRows to obtain S5:

$$S5 = \begin{pmatrix} C11 & C12 & C13 & C14 \\ C23 & C24 & C21 & C22 \\ C31 & C32 & C33 & C34 \\ C43 & C44 & C41 & C42 \end{pmatrix}$$

Then we have,

A) $C11 + C21 + C31 + C41 = 14U1 + 9U2 + 13U3 + 11U4 = T1$

B) $C12 + C22 + C32 + C42 = 11U1 + 14U2 + 9U3 + 13U4 = T2$
C) $C13 + C23 + C33 + C43 = 13U1 + 11U2 + 14U3 + 9U4 = T3$
D) $C14 + C24 + C34 + C44 = 9U1 + 13U2 + 11U3 + 14U4 = T4$

Rewriting this we have.

$$\begin{pmatrix} 14 & 9 & 13 & 11 \\ 11 & 14 & 9 & 13 \\ 13 & 11 & 14 & 9 \\ 9 & 13 & 11 & 14 \end{pmatrix} \begin{pmatrix} U1 \\ U2 \\ U3 \\ U4 \end{pmatrix} = \begin{pmatrix} T1 \\ T2 \\ T3 \\ T4 \end{pmatrix}$$

Proof:

Consider the sum of $T1 = C11 + C21 + C31 + C41$ in S6. We have,

$C11 = 14D11 + 11D24 + 13D33 + 9D42$

$C21 = 9D12 + 14D21 + 11D34 + 13D43$

$C31 = 13D13 + 9D22 + 14D31 + 11D44$

$C41 = 11D14 + 13D23 + 9D32 + 14D41$

Then we have,

$T1 = 14(D11 + D21 + D31 + D41) + 11(D14 + D24 + D34 + D44)$
$\qquad + 13(D13 + D23 + D33 + D43) + 9(D12 + D22 + D32 + D42)$

Therefore, since

$D11 + D21 + D31 + D41 = U1$

$D12 + D22 + D32 + D42 = U2$

$D13 + D23 + D33 + D43 = U3$

$D14 + D24 + D34 + D44 = U4$

we have

$T1 = C11 + C21 + C31 + C41 = 14U1 + 9U2 + 13U3 + 11U4$

Then we have shown condition A) of Property 4b, results for the conditions B), C) and D) can be found in a similar fashion.

QED

## Effect of ShiftRows followed by MixColumns Applied to State S7

Property 5a (Four Shifts):

Let the 128-bit state of AES be S7 from Property 4a, which is such that each of the four shifted 32-bit columns belong to E(U1), E(U2), E(U3) and E(U4) respectively:

$$S7 = \begin{pmatrix} D11 & D12 & D13 & D14 \\ D24 & D21 & D22 & D23 \\ D33 & D34 & D31 & D32 \\ D42 & D43 & D44 & D41 \end{pmatrix}$$

That is, the XOR of each MIxColumns(ShiftRows(S5)) shifted columns are U1, U2, U3 and U4 respectively. We map the 128-bit state down to the 32-bit state vector (U1, U2, U3, U4). S7 is then shifted by ShiftRows once to get the state S8:

$$S8 = \begin{pmatrix} D11 & D12 & D13 & D14 \\ D21 & D22 & D23 & D24 \\ D31 & D32 & D33 & D34 \\ D41 & D42 & D43 & D44 \end{pmatrix}$$

then we apply MixColumns to each column of S8 to get the state S9:

$$S9 = \begin{pmatrix} E11 & E12 & E13 & E14 \\ E21 & E22 & E23 & E24 \\ E31 & E32 & E33 & E34 \\ E41 & E42 & E43 & E44 \end{pmatrix}$$

Then we have,

A) $E11 + E21 + E31 + E41 = U1 = V1$
B) $E12 + E22 + E32 + E42 = U2 = V2$
C) $E13 + E23 + E33 + E43 = U3 = V3$
D) $E14 + E24 + E34 + E44 = U4 = V4$

Proof:

Consider the sum of $V1 = E11 + E21 + E31 + E41$ in S9. We have,

$E11 = 2D11 + 3D21 + D31 + D41$

$E21 = D11 + 2D21 + 3D31 + D41$

$E31 = D11 + D21 + 2D31 + 3D41$

$E41 = 3D11 + D21 + D31 + 2D41$

Then we have,

$V1 = 2(D11 + D21 + D31 + D41) + 3(D11 + D21 + D31 + D41) + (D11 + D21 + D31 + D41) + (D11 + D21 + D31 + D41)$

Therefore, since

$D11 + D21 + D31 + D41 = U1$

we have

$V1 = 2U1 + 3U1 + U1 + U1 = U1$

Then we have shown condition A) of Property 5. The results for the conditions B), C) and D) can be found in a similar fashion.

QED

# Effect of InvMixColumns followed by ShiftRows Applied to State S9

Property 5b (Four Shifts):

Let the 128-bit state of AES be S9 from Property 5a, which is such that each of the four 32-bit columns belong to E(V1), E(V2), E(V3) and E(V4) respectively:

$$S9 = \begin{pmatrix} E11 & E12 & E13 & E14 \\ E21 & E22 & E23 & E24 \\ E31 & E32 & E33 & E34 \\ E41 & E42 & E43 & E44 \end{pmatrix}$$

That is, the XOR of each column of S7 is V1, V2, V3 and V4 respectively. We map the 128-bit state down to the 32-bit state vector (V1, V2, V3, V4). S9 then has InvMixColumns applied to it to get S8:

:

$$S8 = \begin{pmatrix} D11 & D12 & D13 & D14 \\ D21 & D22 & D23 & D24 \\ D31 & D32 & D33 & D34 \\ D41 & D42 & D43 & D44 \end{pmatrix}$$

then, we apply InvShiftRows to obtain S7:

$$S7 = \begin{pmatrix} D11 & D12 & D13 & D14 \\ D24 & D21 & D22 & D23 \\ D33 & D34 & D31 & D32 \\ D42 & D43 & D44 & D41 \end{pmatrix}$$

Then we have,

A) $D11 + D21 + D31 + D41 = U1 = V1$
B) $D12 + D22 + D32 + D42 = U2 = V2$
C) $D13 + D23 + D33 + D43 = U3 = V3$
D) $D14 + D24 + D34 + D44 = U4 = V4$

Proof:

Consider the sum of $U1 = D11 + D21 + D31 + D41$ in S8. We have,

$D11 = 2E11 + 3E21 + E31 + E41$

$D21 = E11 + 2E21 + 3E31 + E41$

$D31 = E11 + E21 + 2E31 + 3E41$

$D41 = 3E11 + E21 + E31 + 2E41$

Then we have,

$U1 = 2(E11 + E21 + E31 + E41) + 3(E11 + E21 + E31 + E41) + (E11 + E21 + E31 + E41) + (E11 + E21 + E31 + E41)$

Therefore, since

$$E11 + E21 + E31 + E41 = V1$$

we have

$$U1 = 2V1 + 3V1 + V1 + V1 = V1$$

Then we have shown condition A) of Property 5. The results for the conditions B), C) and D) can be found in a similar fashion.

QED

## The Effect of SubBytes on Equivalence Classes

In order to identify the effect of SubBytes on Equivalence Classes, we exhaust over a 4-byte state (X1, X2, X3, X4) which are the inputs to SubBytes, compute the corresponding input equivalence class value X = X1+X2+X3+X4 and output equivalence class value Y=Y1+Y2+Y3+Y4 where Yi=SubBytes(Xi), where + is the XOR operation. We compute the counts matrix P_counts[X, Y] and corresponding probability matrix P_dist[X,Y]. In addition, we compute Zi = InvSubBytes(Xi) and compute the counts matrix InvP_counts[X,Y] and corresponding probability matrix InvP_dist[X,Y]. We identify the relationship between P_counts and InvP_Counts.

Some pseudo code is given below.

Compute counts()

int  P_counts[256,256]=0

float P_dist[256,256]=0

int InvP_counts[256,256]=0

float InvP_dist[256.256]=0

for X1=0:255 {

    for X2=0:255 {

        for X3=0:255 {

            for X4=0:255 {

                Y1=SubBytes(X1), Z1=InvSubBytes(X1)

                Y2=SubBytes(X2), Z2=InvSubBytes(X2)

                Y3=SubBytes(X3), Z3=InvSubBytes (X3)

                Y4=SubBytes(X4), Z4=InvSubBytes(X4)

                X=X1+X2+X3+X4

                Y=Y1+Y2+Y3+Y4

$$Z=Z1+Z2+Z3+Z4$$

$$P\_counts[X,Y]++$$

$$InvP\_counts[X,Z]++$$

}}}}

Each cell of P_counts has an expected value of $\frac{2^{32}}{2^{16}} = 2^{16}$ however it turns out that P_counts has an interesting non-uniform distribution. What are the maximum and minimum values of each row?

Row 0 has a maximum value at P_counts[0,0] =198,136 i.e. the mapping of equivalence class value X=0 to equivalence class value Y=0 under SubBytes. The maximum value of every other row is only 68392. The minimum value or Row 0 is 65016, whereas the minimum value of every other row is 64128. This shows that the equivalence class E(0) on input to SubBytes is approximately 3 times more likely than random to map to E(0) on output. This property is very interesting and is quite unexpected.

Similar results can be obtained when using InvSubBytes to obtain InvP_counts.

The relationship between P_counts and InvP_counts is :

P_counts = Transpose (InvP_counts)

which can be shown experimentally.

We have

$$P\_dist[X,Y] = \frac{P\_counts[X,Y]}{2^{24}}$$

and

$$InvP\_dist[X,Z] = \frac{InvP\_counts[X,Z]}{2^{24}}$$

This shows that, in the encrypt direction, if you know the equivalence class value upon input to SubBytes, there is a probability distribution on the equivalence classes on output. Similarly, in the decrypt direction, if you know the equivalence class values upon input to InvSubBytes, there is a probability distribution on the equivalence classes on output.

However, in the encrypt direction, if you know the equivalence class on input to SubBytes and , in the decrypt direction, you know the equivalence class on input to InvSubBytes, then combining these two pieces of information eliminates the probability distributions on equivalence classes.

## Equivalence Classes and the Key Schedule

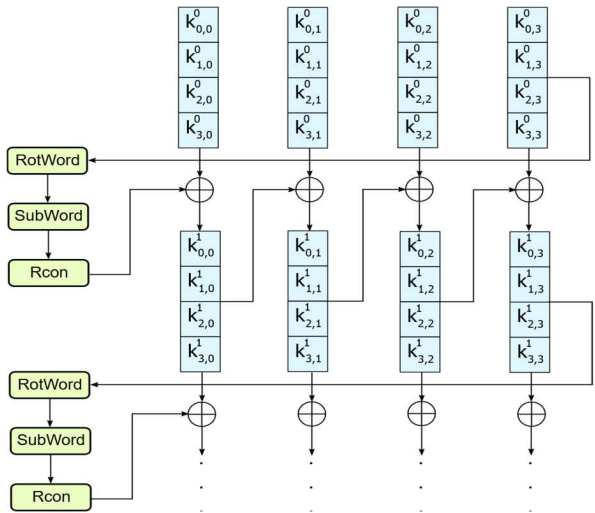

Figure 1: AES-128 Key Schedule (courtesy Wikipedia)

AES-128 has 10 rounds, and therefore a round-key is XOR'd into the state at the end of each of the 10 rounds using AddRoundKey. However, there is an initial XOR of round-key prior to entering round 1. Therefore, there are 11 round-keys in total. We denote the initial round-key as round 0 key and the rest as round 1 key to round 10 key.

The algorithm to generate the keys from round i to round i+1 is as follows:

Let the key at round i be:

$$K(i) = \begin{pmatrix} K00^i & K01^i & K02^i & K03^i \\ K10^i & K11^i & K12^i & K13^i \\ K20^i & K21^i & K22^i & K23^i \\ K30^i & K31^i & K32^i & K33^i \end{pmatrix}$$

Further, let the equivalence classes of the columns of K(i) be denoted as $EK0^i, EK1^i, EK2^i, EK3^i$ which are the respective XOR's of the corresponding columns 0, 1, 2 and 3.

At round i, compute the non-linear function G of the last column of K(i)

$$G(i) = \begin{pmatrix} G0^i \\ G1^i \\ G2^i \\ G3^i \end{pmatrix} = G\begin{pmatrix} K03^i \\ K13^i \\ K23^i \\ K33^i \end{pmatrix}$$

using RotWord, SubWord and RCon. Let the equivalence class of G(i) be denoted as $EG^i$.

Then the process to calculate K(i+1) key from K(i) is:

$$Col\ 0\ of\ K(i+1) = \begin{pmatrix} K00^{i+1} \\ K10^{i+1} \\ K20^{i+1} \\ K30^{i+1} \end{pmatrix} = \begin{pmatrix} K00^i + g0^i \\ K10^i + g1^i \\ K20^i + g2^i \\ K30^i + g3^i \end{pmatrix}$$

$$Col\ 1\ of\ K(i+1) = \begin{pmatrix} K01^{i+1} \\ K11^{i+1} \\ K21^{i+1} \\ K31^{i+1} \end{pmatrix} = \begin{pmatrix} K00^{i+1} + K01^i \\ K10^{i+1} + K11^i \\ K20^{i+1} + K21^i \\ K30^{i+1} + K31^i \end{pmatrix}$$

$$Col\ 2\ of\ K(i+1) = \begin{pmatrix} K02^{i+1} \\ K12^{i+1} \\ K22^{i+1} \\ K32^{i+1} \end{pmatrix} = \begin{pmatrix} K01^{i+1} + K02^i \\ K11^{i+1} + K12^i \\ K21^{i+1} + K22^i \\ K31^{i+1} + K32^i \end{pmatrix}$$

$$Col\ 3\ of\ K(i+1) = \begin{pmatrix} K03^{i+1} \\ K13^{i+1} \\ K23^{i+1} \\ K33^{i+1} \end{pmatrix} = \begin{pmatrix} K02^{i+1} + K03^i \\ K12^{i+1} + K13^i \\ K22^{i+1} + K23^i \\ K32^{i+1} + K33^i \end{pmatrix}$$

Then we have computed:

$$K(i+1) = \begin{pmatrix} K00^{i+1} & K01^{i+1} & K02^{i+1} & K03^{i+1} \\ K10^{i+1} & K11^{i+2} & K12^{i+1} & K13^{i+1} \\ K20^{i+1} & K21^{i+1} & K22^{i+1} & K23^{i+1} \\ K30^{i+1} & K31^{i+1} & K32^{i+1} & K33^{i+1} \end{pmatrix}$$

How do we compute the equivalence classes of EK(i+1) from EK(i)?

We have,

$$EK0^{i+1} = EK0^i + EG(i)$$

$$EK1^{i+1} = EK0^{i+1} + EK1^i$$

$$EK2^{i+1} = EK1^{i+1} + EK2^i$$

$$EK3^{i+1} = EK2^{i+1} + EK3^i$$

# Conclusion

We have defined equivalence classes of AES that arise naturally from properties of MixColumns and InvMixColumns. We have identified the action of SubBytes, ShiftRows, MixColumns and AddRoundKey on equivalence classes.

The next phase of research is to investigate key recovery attacks using equivalence class pairs of (plaintext, ciphertext). The results of this research will be in the Part 2 paper.

# References

1. *Xin An, Kai Hu, Meiqin Wang.* MixColumns Coefficient Property and Security of the AES with A Secret S-Box. 2020/546 Cryptology ePrint Archive
2. *Michael Backes, Lucjan Hanzlik, Kamil Kluczniak, Jonas Schneider.* Signatures with Flexible Public Key: Introducing Equivalence Classes for Public Keys. 2018/191 Cryptology ePrint Archive
3. *Zhenzhen Bao, Jian Guo, Eik List.* Extended Truncated-differential Distinguishers on Round-reduced AES. 2019/622 Cryptology ePrint Archive

4. *Navid Ghaedi Bardeh, Vincent Rijmen.* New Key-Recovery Attack on Reduced-Round AES. 2022/487 Cryptology ePrint Archive
5. *Navid Ghaedi Bardeh.* A Key-Independent Distinguisher for 6-round AES in an Adaptive Setting. 2019/945 Cryptology ePrint Archive
6. *Navid Ghaedi Bardeh, Sondre Rønjom.* Practical Attacks on Reduced-Round AES. 2019/770 Cryptology ePrint Archive
7. *Navid Ghaedi Bardeh, Sondre Rønjom.* The Exchange Attack: How to Distinguish Six Rounds of AES with $2^{88.2}$ chosen plaintexts. 2019/652 Cryptology ePrint Archive
8. *Achiya Bar-On, Orr Dunkelman, Nathan Keller, Eyal Ronen, Adi Shamir.* Improved Key Recovery Attacks on Reduced-Round AES with Practical Data an d Memory Complexities. 2018/527 Cryptology ePrint Archive
9. *Augustin Bariant, Gaëtan Leurent.* Truncated Boomerang Attacks and Application to AES-based Ciphers. 2022/701 Cryptology Archive.
10. *Balthazar Bauer, Georg Fuchsbauer, Fabian Regen*. On Security Proofs of Existing Equivalence Class Signature Schemes. 2024/183 Cryptology ePrint Archive
11. *Aisling Connolly, Pascal Lafourcade, Octavio Perez Kempner.* Improved Constructions of Anonymous Credentials From Structure-Preserving Signatures on Equivalence Classes. 2021/1680 Cryptology ePrint Archive
12. *Joan Daemen and Vincent Rijmen*. The Design of Rijndael: AES - The Advanced Encryption Standard. Information Security and Cryptography. Springer, 2002.
13. *Georg Fuchsbauer, Romain Gay.* Weakly Secure Equivalence-Class Signatures from Standard Assumptions. 2018/037 Cryptology ePrint Archive.
14. *Georg Fuchsbauer, Christian Hanser, Daniel Slamani.* Structure-Preserving Signatures on Equivalence Classes and Constant-Size Anonymous Credentials. 2014/944 Cryptology ePrint Archive.
15. *Steven D. Galbraith, Raminder S. Ruprai.* Using Equivalence Classes to Accelerate Solving the Discrete Logarithm Problem in a Short Interval. 2010/615 Cryptology ePrint Archive
16. *Agnese Gini, Pierrick Méaux.* S0-equivalent classes, a new direction to find better weightwise perfectly balanced functions, and more. 2023/1101 Cryptology ePrint Archive
17. *Lorenzo Grassi.* Mixture Differential Cryptanalysis and Structural Truncated Differential Attacks on round-reduced AES. 2017/832 Cryptology ePrint Archive
18. *Lorenzo Grassi, Gregor Leander, Christian Rechberger, Cihangir Tezcan, Friedrich Wiemer.* Weak-Key Distinguishers for AES. 2019/852 Cryptology ePrint Archive.
19. *Lorenzo Grassi, Markus Schofnegger.* Mixture Integral Attacks on Reduced-Round AES with a Known/Secret S-Box. 2019/772 Cryptology ePrint Archive
20. *Lorenzo Grassi, Christian Rechberger.* Truncated Differential Properties of the Diagonal Set of Inputs for 5-round AES. 2018/182 Cryptology ePrint Archive
21. *Christian Hanser, Daniel Slamanig.* On Enumeration of Polynomial Equivalence Classes and Their Application to MPKC. 2011/055 Cryptology ePrint Archive
22. *Kai Hu, Tingting Cui, Chao Gao, Meiqin Wang.* Towards Key-Dependent Integral and Impossible Differential Distinguishers on 5-Round AES. 2018/726 Cryptology ePrint Archive
23. *Seonggyeom Kim, Deukjo Hong, Jaechul Sung, Seokhie Hong.* Accelerating the Best Trail Search on AES-Like Ciphers. 2022/643 Cryptology ePrint Archive

24. *Mojtaba Khalili, Daniel Slamanig, Mohammad Dakhilalian.* Structure-Preserving Signatures on Equivalence Classes From Standard Assumptions. 2019/1120 Cryptology ePrint Archive.
25. *Gaëtan Leurent, Clara Pernot.* New Representations of the AES Key Schedule. 2020/1253 Cryptology ePrint Archive.
26. *Omid Mir, Daniel Slamanig, Balthazar Bauer, René Mayrhofer.* Practical Delegatable Anonymous Credentials From Equivalence Class Signatures. 2022/680 Cryptology ePrint Archive
27. *Yanbin Pan.* Cryptanalysis of the Structure-Preserving Signature Scheme on Equivalence Classes from Asiacrypt 2014. 2014/915 Cryptology ePrint Archive.
28. *Bing Sun, Meicheng Liu, Jian Guo, Longjiang Qu, Vincent Rijmen.* New Insights on AES-like SPN Ciphers. 2016/533 Cryptology ePrint Archive
29. *Aurelien Vasselle, Antoine Wurcker.* Optimizations of Side-Channel Attack on AES MixColumns Using Chosen Input. 2019/343 Cryptology ePrint Archive